# Generating Music using an LSTM Network

Nikhil Kotecha nsk2147,  Paul Young py2227
*Columbia University*

## Abstract

*A model of music needs to have the ability to recall past details and have a clear, coherent understanding of musical structure. Detailed in the paper is a neural network architecture that predicts and generates polyphonic music aligned with musical rules. The probabilistic model presented is a Bi-axial LSTM trained with a "kernel" reminiscent of a convolutional kernel. When analyzed quantitatively and qualitatively, this approach performs well in composing polyphonic music. Link to the code is provided[1].*

## 1. Introduction

This paper describes an algorithmic approach to the generation of music. The key goal is to model and learn musical styles, then generate new musical content. This is challenging to model because it requires the function to be able to recall past information to project in the future. Further, the model has to learn the original subject and transform it. This is a non-trivial task. The next challenge is to understand the underlying substructure of the piece so that it performs the piece cohesively. It is easier to create small, non-connected subunits that do not contribute to sense of a coherent piece.

One method is to train a probabilistic model of music. Model the music as a probability distribution, mapping measures, or sequences of notes based on likelihood of appearance in the corpus of training music. These probabilities are learnt from the input data without prior specification of particular musical rules. The algorithm uncovers patterns from the music alone. After the model is trained, new music is generated. This generated music comes from a sampling of the learned probability distribution. This approach is complicated by the structure of music. Structurally, most music contains a melody, or a key sequence of notes with a single instrument or vocal theme. This melody can be monodic, meaning at most one note per time step. The melody can also be polyphonic, meaning greater than one note per time step[2]. In the case of Bach's chorales, they have a polyphony, or multiple voices producing a polyphonic melody. These melodies can also have an accompaniment. This can be counterpoint, composed of one or more melodies or voices[3]. A form of accompaniment can also be a sequence of chords that provide an associated term called a harmony. The input has great bearing on the nature of the output generated.

These musical details are relevant because training a probabilistic model is complicated by the multidimensionality of polyphonic music. For instance, within a single time step multiple notes can occur creating harmonic intervals. These notes can also be patterns across multiple time steps in sequence. Further, musical notes are expressed by octave, or by interval between musical pitches. Pitches one or more octaves apart are by assumption musically equivalent, creating the idea of pitch circularity. Pitch is therefore viewed as having two dimensions: height, which refers to the absolute physical frequency of the note (e.g. 440 Hz); and pitch class, which refers to relative position within the octave. Therefore, when music is moved up or down a key the absolute frequency of the note is different but the fundamental linkages between notes is preserved. This is a necessary feature of a model. Chen et al[4] offered an early paper on deep learning generated music with a limited macro structure to the entire piece. The model created small, non-connected subunits that did not contribute to a sense of a coherent composition. To effectively model music, attention needs to be paid to the structure of the music.

A model of music needs to have the ability to recall past details and understand the underlying sub-structure to create a coherent piece in line with musical structure. Recurrent neural networks (RNN), and in particular long short-term memory networks (LSTM), are successful in capturing patterns occurring over time. To capture the complexity of musical structure vis a vis harmonic and melodic structure, notes at each time step should be modeled as a joint probability distribution. To account for octaves and pitch circularity, greater context is needed. Following the convolutional neural network architecture, a solution is to employ a kernel or a window of notes and sliding that kernel or convolving across surrounding notes. Inspired by Daniel Johnson's[5] Bi-axial LSTM model, we describe a neural network architecture that generates music. The probabilistic model described is a stacked recurrent network with a structure employing a convolution-esque kernel. Presented is model of the original paper, our changes to the model, our approach to training, and generation. Our code is available [1].

## 2. Methodology

In this section, presented is Daniel's Johnson's original model followed by our extensions to the model. In the original paper there are a few models attempted to generate music. Here we select the best performing model, replicate, and extend the model.

## 2.1. Objectives and Technical Challenges

One key challenge with modeling music is selecting the data representation. Possible representations are signal, transformed signal, MIDI, text, etc. A relevant issue is the end destination of the generated music content[6]. The format destination could be a human user, in which case the output would need to be human readable, for instance a musical score. In the case of this paper, the destination is a computer. The final output format is therefore readable by a computer, which in this case is a MIDI file (musical instrument digital interface). The MIDI representation was selected because it offers a particularly rich representation in two senses: first it carries characteristics of the music in the metadata of the file, like time steps. Second it is a common digital representation which allowed access to freely and widely available data.

Another relevant factor is the level of supervision in the generation of the output. At one extreme is complete autonomy and automation with no human supervision. Or it could be more interactive, with early stopping built into the model to supervise the music creation process. The neural network approach employed by this paper is by design non-interactive. The MIDI file format optimized for this dimension as well because it offers a complete end product that is machine readable without human intervention. The level of autonomy is an interesting potential development for actual musicians who can interrupt the model in the middle of content generation. While beyond the scope of the paper, feedback throughout the process can lead to suggestions that are superior, or more aesthetically pleasing musical compositions.

## 2.2. Problem Formulation and Design

Recurrent networks encounter a serious problem caused by difficulty in estimating gradients. In backpropagation through time (BPTT), recurrence passes multiplications in repetition. This can lead to diminishingly small or increasingly large effects, respectively called the vanishing or exploding gradient problem. To resolve this problem, Hochreiter and Schmidhuber[7] designed Long short-term memory (LSTM) networks. The LSTM is designed to secure information in memory cells, separate and protected from the standard information flow of a recurrent network. To pass, read or forget information is performed by opening or closing the gates, akin to a neuron firing. This is learned during the training process. Gates are modulated by weight that is differentiable, allowing for back propagation in typical neural network learning fashion.

To capture the harmonic and melodic structure between notes, the model uses a two-layered LSTM RNN architecture with recurrent connections along the note axis. By having one LSTM on the time axis and another on the note axis, the model takes on, to borrow Daniel Johnson's language[5]: a "bi-axial" configuration.

The note-axis LSTM receives as input a concatenation of final output of the note-axis LSTM for the previous note window and the activations of the last time-axis LSTM layer for the particular note. The output of the final activations of the note-axis LSTM are then fed into a softmax layer to convert to a probability. The loss corresponds to the cross entropy error of the predictions at each time step compared to the played note at each time step. Each note therefore has a time component from the time-axis LSTM. This allows for understanding the temporal relationships for the particular note and for modeling the joint distribution of notes in the particular time step. By joining the information from an LSTM focused on the time component and an LSTM focused on the note-component, the relationships within and between notes is captured for each timestep. By using this approach on each note in sequence, the full conditional distribution for each time step can be learned. Further, another key piece of functionality is building into the model a window that slides over sequences of notes. This architecture enables the model to learn the harmonic and melodic structure of the notes accounting for pitch circularity.

Extending beyond Daniel Johnson's model, the model presented here is designed and implemented to be flexible, general, and to take advantage of parallelization in code. A primary goal was flexibility in user input. Our architecture is general: the user can set various hyper parameters, such as the number of layers, hidden unit size, sequence length, time steps, batch size, optimization method, and learning rate. The model is parameterized so users can also set the length of the window of notes fed into the note-axis LSTM model and the length of time steps fed into the time-axis LSTM. The size of the window and length of the time steps are a relevant features because music is highly variable based on genre and artist. Designing the system to be general allows the user to tailor the model to his/her specific needs. In terms of functionality and model design, a primary goal for the model was parallelization in code. The code was written at a high level to do everything in efficient matrix format, minimizing the use of 'for' loops. This allows for speed gains in computational time.

## 3. Implementation

In this section, the process of training the network and the generation of new musical compositions will be explained. Experiments were performed on Google Cloud Platform with deep learning implementation done in TensorFlow. Sources of material that helped guide the implementation: Daniel Johnson's code [8]. For loading the data into the appropriate format [9].

### 3.1. Deep Learning Network

The model is applied to a polyphonic music prediction task. The network is trained to model the conditional probability distribution of the notes played in a given time step, conditioned on the notes in previous time steps. The output of the network can be read as at time step t, the probability of playing a note at time step t, conditioned on prior note choices. Therefore, the model is maximizing the log-likelihood of each training sequence under the conditional distribution.

The time-axis LSTM depends on chosen notes, not on the specific output of the note axis layers. The rationale is that all notes at all timesteps are known so training can be expedited. The time gain comes from processing the input, then feeding the pre-processed input through the LSTM time-axis in parallel for all notes. Next, the LSTM note-axis layer computes the probabilities across all time steps. This provides a significant speed up when using a GPU to perform parallel computing.

Now that the probability distribution is learned, sampling from this distribution offers a way to generate new sequences. Sequences are not known in advance. The network must project one time step in the future at a time. The input for each timestep is used to advance the LSTM time-axis layers one step at at a time to compose the note in the next period. First a sample must be taken while the distribution is being created. Each note is drawn from a Bernoulli distribution. This drawn value is then used for the input to the next note. This process is repeated for all notes, after which the model moves to the next time step.

The model was tested on a selection of Bach's works from [17] as well as the classical piano files from [18].. Input was in the form of MIDI files.

After training the Bi-axial LSTM, the model was used to create new musical compositions. A larger and diverse dataset with different note and structural patterns was used during training. The goal here was to expose the model during training to a wide variety of patterns so as to encourage as much diversity in output as possible. The MIDI file format enables the use of a temporal position in the music. A time component was an important feature to build into the dataset so that the model could learn patterns over time relative to different note sequences. Following the guide of Johnson[5], an additional dimension was added to the note vectors fed into the model: a binary, 0 or 1 to indicate if a note was articulated or sustained at a particular time step. From Johnson, for instance, the first time step for playing a note is represented as 11. Sustaining a previous note is represented as 10, and resting is represented as 00. This added dimension allows the model to play the same note multiple times in succession. From the input perspective, the articulation dimension or bit is processed beforehand. This processing is done in parallel with the playing dimension, which together are then fed into to the time-axis LSTM. From the output perspective, the note-axis LSTM gives a probability of playing a note and a probability of articulating the same note. When computing the cost function, articulating a played note incorrectly is penalized. The articulation output for notes that should not be played is ignored. It makes little sense to penalize for articulation if a note is not played.

Using Moon et al[14] as a suggested guide, Dropout of .75 was applied to each LSTM layer. The optimizer selected was ADADELTA[15]. The learning rate selected was 1.0. The models were evaluated in two dimensions.

### 3.2. Software Design

The featured data vector in this neural network is referred to as the 'Note State Matrix' shown in Figure 1. This represents the 'play' and 'articulate' state of each note over the range of Midi values and for each time step over a specified period of time (i.e. 8 measures at 16 time steps per measure). The model takes as input a single batch of these feature data vectors, a single 4D tensor referred to as the 'Note_State_Batch'. The original raw musical data in the form of .MIDI files are first preprocessed to generate each Note_State_Batch using the Python-Midi package extracted from [16]. In this work, we used this package only to import MIDI file segments as Note_State_Batches, as well as to create MIDI files from our Generated Samples. Since these are common pre/post processing tasks, it was deemed of little value for the purposes of this neural networks and deep learning class to recreate them. However, it may be of interest in further work to enhance the processed feature data vectors to include other musical features such as volume.

$$\textit{Note State Matrix} = \begin{bmatrix} [p,a]_N^{(1)} & \cdots & [p,a]_N^{(T)} \\ & \vdots & \\ [p,a]_1^{(1)} & \cdots & [p,a]_1^{(T)} \end{bmatrix}$$

a batch of which constitute a 'Note_State_Batch'

Fig. 1: Note State Matrix is the processed, feature vector for the neural network. N is the # Midi note states extracted from the songs, T is the # time steps in the batch, 'p' indicates the binary value of the note being played, and 'a' indicates the binary value of the corresponding articulation.

The overall structure of our code was broken into two main tasks: training/validating the model numerically and then using the trained model to generate new .MIDI files for qualitative evaluation. Both functions utilize the same Model Graph in different contexts: The training task, shown at a high level in Figure 2, iteratively inputs a Note_State_Batch into the model, runs the model through all of the corresponding time steps and notes present in the batch, and then outputs a tensor of corresponding 'Logits', or inverse sigmoid probability that a given note at a given time step is played/articulated. The log likelihood of the input data is interpreted as the ability of the model to take as input a vector of notes at a given time step and to predict the set of notes at the subsequent time step. The Loss Function, pseudo-code of which is shown in Figure 3, calculates cross-entropy between the generated Logits and the Note_State_Batch (after lining up the Logits to the Note_State_Batch elements corresponding to one time step in the future).

During the music generation task, presented in Fig. 4, the model is iteratively run through one time step, every time feeding back the Generated Samples as the Note_State_Batch input for the subsequent time step. This samples are accumulate, and this produces a tensor of Generated Samples in the form of the Note_State_Batchof arbitrary time length. The Generated Samples are then converted to .MIDI files using post-processing functions from [16] for qualitative evaluation.

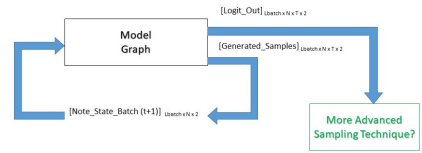

Fig. 4: High-level view of music generation graph. Includes tentative plans for enhancement.

A functional breakdown of the Model Graph, itself, is shown in Figure 5. Pseudo code of the first stage of the model, referred to as the 'Input Kernel'', is shown in Figure 6. The Input Kernel takes a Note_State_Batch as its input and for each note/articulation pair, generates an expanded vector that consists of: 1) the Midi note number, 2) a one hot vector of the note's pitch class, 3) window of the play/articulation values relative to the 'n'th note (the effective convolutional kernel aspect of the model), 4) a vector of the sum of all played notes in each pitch class, and 5) a binary-valued vector representing the 16 value position of the note within a measure.

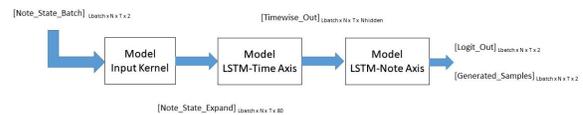

Fig. 5: Breakdown of Model Graph

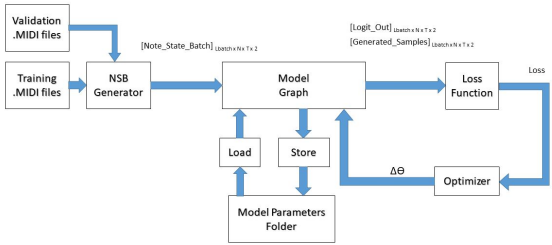

Fig. 2: High-level view of the Training Graph

Arguments:
- [Logits] $L_{batch} \times N \times T \times 2$ (inverse sigmoid of Probability that play/articulate = 1)
- Labels(t-1) = [Note_State_Batch] $L_{batch} \times N \times T \times 2$

$$\text{cross\_entropy} = \text{sigmoid\_cross\_entropy\_with\_logits}(\text{logits}=\text{logits}, \text{labels}=\text{Labels})$$
$$= -\text{note\_state} \cdot \ln(\sigma(\text{logits})) + -(1-\text{note\_state}) \ln(1-\sigma(\text{logits}))$$
$$= -\text{note\_state} \cdot \ln(\text{Probability}=1) + -(1-\text{note\_state}) \ln(\text{Probability}=0)$$

$$\text{Loss} = \frac{1}{TNL_{batch}} \sum_{b=1}^{L_{batch}} \sum_{t=1}^{T} \sum_{n=1}^{N} \text{cross\_entropy}$$

$$\text{Log-likelihood at 1 time step} = -\frac{1}{TL_{batch}} \sum_{b=1}^{L_{batch}} \sum_{t=1}^{T} \sum_{n=1}^{N} \text{cross\_entropy}$$

Returns:
- Loss (scalar)

Fig. 3: Pseudo-code for Loss definition and calculation

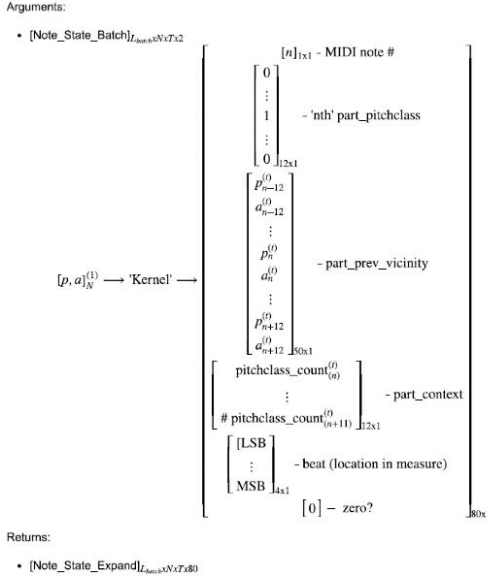

Fig. 6: Pseudo code for Input Kernel. This code is performed in parallel note-wise, time-wise, and sample-wise.

The second stage is referred to as the Timewise LSTM stage, pseudo-code for which is shown in Figure 7. In this block, an LSTM cell is run along the time axis for the length of the batch time dimension. This operation is performed on the Note_State_Expand vector for every note in parallel with tied weights. This part of the graph captures the sequential patterns of the music and, in combination with the Input Kernel, preserves translation invariance due to the input window of relative notes and the tied LSTM weights across all notes. Due to these tied weights, the computations can be run in parallel across notes and across Note State Matrix samples as separate effective batches. The only required sequential aspect is along the time axis. An arbitrary number of cascaded LSTM cells can be run, and a dropout mask is applied after each cell.

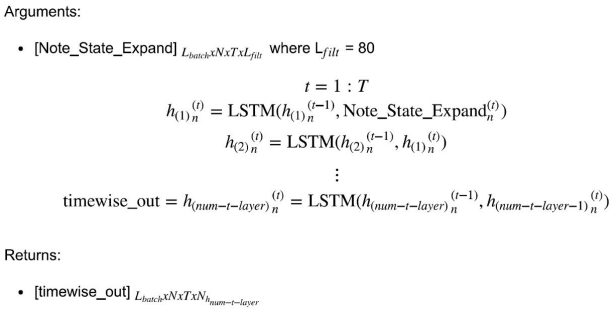

Fig. 7: Pseudo code for Timewise LSTM stage. This code is performed in parallel note-wise and sample-wise

The final stage in the Model Graph as described in the block diagram is the Notewise LSTM stage, pseudo-code for which is shown in Figure 8. This is a potentially one or multi-layered LSTM stage like the Timewise LSTM, also with dropout after each layer. However, instead of running sequentially along the time axis, this stage runs sequentially along the note axis. Furthermore, this section includes the 'local' feedback of generated samples into its input. After each 'note step', the LSTM cell produces a pair of logits representing the inverse sigmoid of the probability of generating a play/articulation for that note. Next, a play and articulation sample are drawn from this Bernoulli distribution. If the play sample is a '0' for 'not played', the articulation sample is forced to '0', as well, to avoid the generation of any values not present in the input data. The generated sampled pair at note (n-1), concatenated with the input of the timewise LSTM stage at note (n), is fed back into the input of the notewise LSTM for step (n). This feedback creates a conditional probability for each note based on the actual values generated for lower notes. This helps prevent dissonant simultaneous notes from being played. The final output tensors of the Model Graph are the batch of Logits and corresponding Generated Samples that are used for training and music generation, respectively.

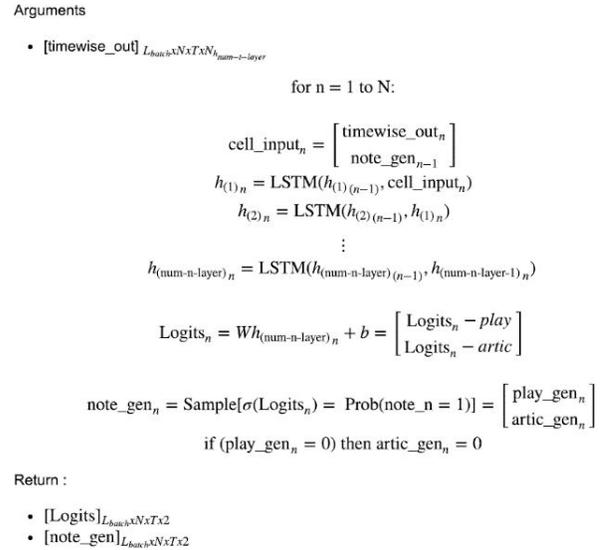

Fig. 8: Pseudo Code for Notewise LSTM Stage. This code is performed in parallel time-wise and sample-wise.

## 4. Results
### 4.1. Quantitative Analysis

Figure 9 shows the log-likelihood performance of the author's models, as well as that of the model implemented in this work. The results obtained in this work were, in general, on par with the survey of models reported by the original author but somewhat inferior to the corresponding model. However, limitation of training time (solely due to time constraints) most likely plays the largest role in this discrepancy. In his blog, the original author estimated roughly 24-48 hours of training to capture the quality of his music samples, whereas the training for this work consisted of about 13 hours on a limited set of 1-2 dozen of Bach's fugues from [17], this part of the training is displayed in Figure 10, followed by less than 3 hours on the full Piano-Midi.de dataset, shown in Figure 11. This progression of complexity was performed while validating basic model functionality. It is expected that with the model finalized, a much longer training run will be performed to obtain better quantitative results, though this will likely not be reported. In addition, the author's paper reported optimization using RMSprop whereas his blog, which seemed to represent the latest of his progress, reported Adadelta. This work started with the latter, but more experimentation needs to be done to fine tune such hyper parameters.

| Model | Log-Likelihood | Hours Trained |
| --- | --- | --- |
| Random | -61 | -- |
| TP-LSTM-NADE | -5.44, -5.49 | 24 - 48 |
| BALSTM | -4.90, -5.00 | 24 - 48 |
| BALSTM (this work) | -6.27, -7.93 (test) -5.16, -6.59 (train) | 16 |

Fig. 9: The top 3 rows represent the Log-likelihood performance reported by the original author for random weighted, LSTM-NADE, and Bi-axial LSTM networks tested on the Piano-Midi.de data set. The two values represent the best and median performance across 5 trials. For the bottom row, the two values represent best and median across 100 trials for the BALSTM in this work, scaled by 88/78 to normalize it to the number of notes used by the original author.

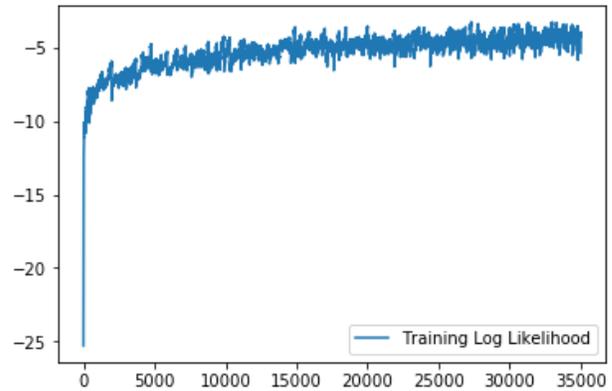

Fig. 10: Pre-training log likelihood for data set consisting of 20 of Bach's fugues. The data had been pre-trained on 4 beginner piano music songs for 4 hours. The 35,000 iterations in this plot took an additional 9 hours.

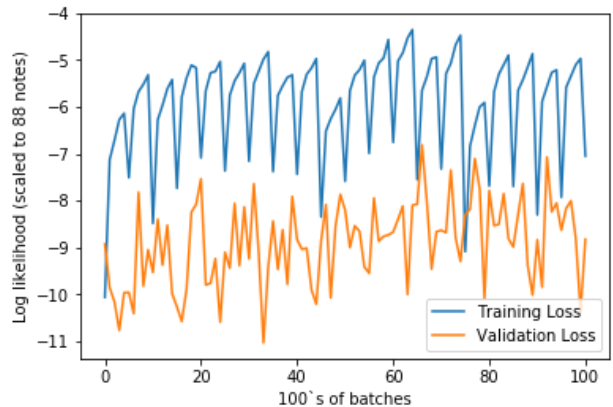

Fig. 11: Final training of our model using Piano-Midi.de dataset for 2 hrs 45 min. This training was a continuation of that in Figure 9. The up and down 'jogs' in the training loss represent new batches being sampled.

### 4.2. Qualitative Analysis

Qualitatively, the samples produced by this model produce decent rhythm and does well at creating small stretches of notes with melody, simple harmony, and in a few cases even counterpoint. These stretches of notes are polyphonic and have a local sense of coherence. The music breaks down in its ability to create clear transitions between larger ideas in the piece as a whole. There is no deeper structure. The sample also makes poor use of negative space, few pauses are present in the work. Due to the lack of global structure, the music has a mechanical feel. An important note is the length of training time. When the model is trained for 30 minutes, the music generated is sparse and significantly less consistent and coherent. When trained for 2 hours, the difference is dramatic. Clear relationships between generated music and the corresponding training files developed.

### 4.3. Discussion of Insights Gained

It became clear how the variability and complexity of music on which the model was trained affected the outcome. Training a newly initialized model on a large data set consisting of significant variability in music segments (i.e. fast monodic and slow polyphonic) tended to create a model that seemed to be confused at first. Trying to learn such a range of features requires a very complex function needing very long training times. Training on a set consisting of 22 of Bach's fugues from [17] obtained better results more quickly than training on the 120 Piando-de-Midi for modest training times < 2 hours. However, it became evident that very long training times were required, in general to produce decent music. It was apparent that the training graphs did not always follow a relatively exponential-like curve. In many cases, the training loss would appear to be settling for 1-2 hours, and then begin to decrease heavily for another couple hours. The quality of the music as training time increased seemed to reflect the quantitative training progress. It was clear the music was gradually learning rhythm and chord structures, however it sounded as if a human were learning to play piano but trying to play songs that were too difficult. One possible training strategy may be to train on a succession of increasingly difficult songs, graduating the model manually, or perhaps in an automated fashion once a certain ability/log likelihood was achieved. In addition to songs of different 'level-of-difficulty', training could begin on very short time segments and increase to very long segments to allow the model to learn basic structure in addition to longer musical form.

In terms of future work, it would be fruitful to add to the bi-axial LSTM a component that focused on structure alone. There has been good work showing the merits of using Restricted Boltzmann Machines to model chord progressions and other forms of harmonic and melodic structure. Additionally, an effective model could incorporate genetic algorithms. The line of thinking would be to train the model on some simple music and set the fitness score as a proxy for novelty, and allow the algorithm to generate mutations to add complexity to the piece over time. Another model design that would be effective would be Generative Adversarial Networks (GANs) which have achieved remarkable progress in generating photo-realistic images and as such should provide effective musical generation. A more innovative approach is to rely on reinforcement learning and incorporate a sense of exploration in the music generation.

The idea is to use the trained LSTM and tune the hyperparameters with a, for instance, Q learning algorithm[10]. This mechanism works by learning an action value function and following an optimal policy through. A potential refinement can enter in the sampling: paths or musical measures explored may have different rewards associated with the distribution. To effectively explore and learn the potential rewards, a Bayesian updating by sampling different distributions can occur. This exploration or sampling can occur through a naive epsilon greedy mechanism or a upper confidence bound or probably most effectively the Thompson Sampling mechanism [11]. Using a reinforcement learning paradigm coupled with a deep learning technique would allow for effective modeling of the underlying musical structure and for increased range in potential musical expression.

### 5. Conclusion

In this paper, a two layer LSTM model capable of learning harmonic and melodic rhythmic probabilities from polyphonic MIDI files of Bach. The model design was explained, with an eye to key functional principles of flexibility and generalizability. The underlying logic and method of training and generation of algorithmic music were presented. Further, the outputs of the model were analyzed in a quantitative and qualitative fashion. Some suggestions were then put forward for future work.

### 6. Acknowledgement